\documentclass[twocolumn,showpacs,preprintnumbers,amsmath,amssymb,floatfix,nofootinbib]{revtex4-1}
\usepackage{graphicx,color}
\usepackage{dcolumn}
\usepackage{bm}
\usepackage{amsthm}
\usepackage{amsmath}
\usepackage{amsfonts}

\usepackage{mathrsfs}

\newcommand{\n}[1]{\label{#1}}
\newcommand{\eq}[1]{(\ref{#1})}

\newcommand{\be}{\begin{equation}}
\newcommand{\ee}{\end{equation}}
\newcommand{\ba}{\begin{eqnarray}}
\newcommand{\ea}{\end{eqnarray}}

\newcommand{\non}{\nonumber}

\newcommand{\hh}{\hspace{0.2cm}}
\newcommand{\hhh}{\hspace{0.5cm}}

\newcommand{\BM}[1]{{\mbox{\boldmath $#1$}}}
\newcommand{\ind}[1]{\mbox{\tiny $#1$}}

\begin{document}
 
\title{Geometric Properties of Stationary and Axisymmetric Killing Horizons} 

\author{Andrey A. Shoom}
\email{ashoom@ualberta.ca}
\affiliation{Theoretical Physics Institute, University of Alberta, 
Edmonton, AB, Canada,  T6G 2E1}

\begin{abstract}

We study some geometric properties of Killing horizons in 4-dimensional stationary and axisymmetric space-times with electromagnetic field and cosmological constant. Using a $(1+1+2)$ space-time split, we construct relations between the space-time Riemann tensor components and components of the Riemann tensor corresponding to the horizon surface. The Einstein equations allow to derive the space-time scalar curvature invariants, Kretschmann, Chern-Pontryagin, and Euler, on the 2-dimensional spacelike horizon surface. The derived relations generalize the relations known for Killing horizons of static and axisymmetric 4-dimensional space-times. We also present the generalization of Hartle's curvature formula.

 \end{abstract}

\pacs{04.20.-q, 04.20.Cv, 04.70.-s, 04.70.Bw \hfill  
Alberta-Thy-16-14}

\maketitle

\section{INTRODUCTION}

Killing horizons play a significant role in the analysis of pseudo-Riemannian manifolds and are important characteristics of such manifolds. They help to define the global structure of space-time, as black-hole event horizons, Cauchy horizons, cosmological event horizons, and local isometry horizons (for details see \cite{Carter,Carter1,Hawking} and references wherein). A Killing horizon is a null hypersurface in a pseudo-Riemannian manifold which is invariant with respect to a one parameter group of isometries of the manifold and its null geodesic generator is an orbit of the group \cite{Carter}. In a 4-dimensional space-time, a 2-dimensional spacelike Killing horizon surface is a marginally locally trapped surface whose future-directed null normals are not expanding. The generator of a Killing horizon, which is a null Killing vector field, has many interesting geometric properties explored in the works of Carter \cite{Carter}, Boyer \cite{Boyer}, and Wald \cite{Wald2}. The reader can find the comprehensive presentation of many of such properties in the meaty book \cite{FN}. 

Due to special features of a Killing horizon, the corresponding space-time structure takes a special form on and in the vicinity of it. In particular, the space-time geometry and the Einstein equations get simplified due to an enhancement of the space-time symmetries in space-times with the so-called extremal Killing horizon. The well-known example is that of the extreme Kerr black hole solution there the near horizon geometry (the extreme Kerr throat) has enhanced symmetry, and, as a result, the Killing tensor becomes reducible (see, e.g., \cite{BH}). There are many examples of symmetry enhancement of the near horizon geometry of extreme (as well as supersymmetric) horizons in 4 and higher-dimensional space-times (see, e.g., \cite{NH1,NH2,NH3,NH4} and references therein). There are other examples illustrating the special nature of a Killing horizon. It was demonstrated that space-times of local 4-dimensional vacuum black holes represented by static and axisymmetric Weyl solutions of the vacuum Einstein equations are of the Petrov type I (algebraically general), but they become of Petrov type D on the horizon due to the ``appearance'' of two repeated principal null directions \cite{PX}. The same situation takes place for the inner and outer horizons of a local 4-dimensional static and axisymmetric electrically charged black hole \cite{AFS}. It was shown that space-time scalar curvature invariants get greatly simplified when calculated on a Killing horizon (see, e.g., \cite{FS,AFS,AS}).  

In this paper we shall study geometric properties of Killing horizons in 4-dimensional stationary and axisymmetric space-times with electromagnetic field and cosmological constant. We shall not be interested in the global space-time structure and shall study Killing horizons quasilocally. In this sense, the Killing horizon is a particular class of the so-called isolated horizons, which were defined and later extensively studied in e.g., \cite{Ash,Ivan1,Ivan2,Ivan3,Ivan4,Ivan5,Ivan6}. We shall focus on space-time curvature invariants calculated on a Killing horizon. There are 14 algebraically independent scalar invariants constructed from the Riemann curvature tensor \cite{Witten}. Note that a space-time metric of a 4-dimensional Lorentzian manifold can be completely characterized by scalar polynomial curvature invariants constructed from the Riemann tensor and its covariant derivatives, except for the case when its metric is of degenerate Kundt form \cite{Alan}. Here, we will calculate the second order space-time scalar curvature invariants, the Kretschmann, Chern-Pontryagin, and Euler invariants (see, e.g., \cite{Ruf}) on a stationary Killing horizon. Killing horizons considered in this paper are regular in the sense that these invariants are finite. The results derived here is an extension of the previous works  \cite{FS,AFS,AS} where the Kretschmann invariant was calculated on static Killing horizons. The Kretschmann scalar of a Killing horizon in a 4-dimensional electrovacuum (without magnetic field) static space-time was derived in \cite{AFS},
\ba\n{I1}
{\cal K}\circeq3\left({\cal R}+F^2\right)^2+2F^4\,,
\ea
where ${\cal R}$ is the Ricci scalar of the horizon 2-dimensional spacelike surface and $F^{2}=F_{\alpha\beta}F^{\alpha\beta}$ is the electromagnetic field invariant.\footnote{In this paper we use the symbol $\circeq$ to define a relation between quantities calculated on a Killing horizon.} Another work \cite{AS} contains a study of Killing horizons within the $d$-dimensional Einstein-Maxwell-dilaton model with a cosmological constant.  

Beside an analysis of geometric properties of a Killing horizons, the sought relations have many applications. For instance, the expression of the Kretschmann scalar was used in \cite{Werner1,Werner2} to prove the uniqueness theorems for the Schwarzschild and Reissener-Nordstr\"om black hole solutions. An investigation of properties of scalars and tensor invariants constructed from the Weyl tensor, Killing vector, and their derivatives near a Killing horizon is necessary to calculate the vacuum energy density near a static $4$-dimensional black hole using Page's \cite{P} and Brown's \cite{B} approximations (see, e.g., \cite{FS}). The relation \eq{I1} was used in \cite{AFS} to analyze the curvature of the (inner) Cauchy horizon of a distorted, static, and axisymmetric Reissner-Nordstr\"om black hole based on the curvature of its outer horizon. Such an analysis was possible due to a certain duality transformations between the black hole's horizons. The relations derived in this paper can help to analyze the curvature of the Cauchy horizon of a distorted, stationary, and axisymmetric Kerr-Newman black hole solution constructed in \cite{BGMD}.

Our paper is organized as follows: In Sec. II we construct the metric of a stationary and axisymmetric space-time in (1+1+2)-split form that allows for the space-time foliation suitable for studying of the Killing horizon surface. In Sec. III we derive relations between the space-time Riemann tensor components and components of the Riemann tensor corresponding to the horizon surface. Section IV contains the Einstein equations of a stationary and axisymmetric space-time with an electromagnetic field (without a source) and a cosmological constant and expressions of the space-time curvature invariants in the form corresponding to the (1+1+2)-split of the metric. In Sec. V we define the Killing horizon and using the results of the previous sections calculate the space-time curvature invariants on the horizon surface. Section VI contains discussion of the derived results and present them in terms of the gravitoelectric and gravitomagnetic fields.

In this paper we use the following convention of units: $G=c=1$. The space-time signature is $+2$, and the sign conventions are that adopted in \cite{MTW}.

\section{Space-time split}

In this section, we construct metric of a 4-dimensional stationary and axisymmetric space-time and present it in $(1+1+2)$-split form.   We consider a 4-dimensional Lorentzian manifold $({\cal M},g_{\alpha\beta})$, where $g_{\alpha\beta}$ satisfies the Einstein equations, which has a two-parameter abelian group of isometries $\{\varphi_{t},\varphi_{\phi}\}$. Orbits of $\varphi_{t}$ are timelike at asymptotic infinity and orbits of $\varphi_{\phi}$ are spacelike and closed. The generators of the group are the commuting Killing vector fields $\BM{\xi}_{(t)}$ and $\BM{\xi}_{(\phi)}$, which are not orthogonal. We choose the space-time coordinates such that  $\xi^{\alpha}_{(t)}=\delta^{\alpha}_{\;t}$, where $t$ is time coordinate and $\xi^{\alpha}_{(\phi)}=\delta^{\alpha}_{\;\phi}$, where $\phi$ is a spatial coordinate, which in the axisymmetric case is an azimuthal angular coordinate. A space-time is called stationary (pseudo-stationary, in the case when the Killing vector field $\BM{\xi}_{(t)}$ is not timelike everywhere), but not static, if the timelike Killing vector $\xi^{\alpha}_{(t)}$ is not hypersurface orthogonal, i.e., the condition 
\be
\xi^{[\alpha}_{(t)}\nabla^{\beta}\xi_{(t)}^{\gamma]}=0\,
\ee
does not hold. Otherwise, it is called static, which is a special case of being stationary. Here and in what follows, the symbol $\nabla$ stands for a covariant derivative defined with respect to the metric $g_{\alpha\beta}$. 

Let us now consider a hypersurface $\Sigma_{t}$ defined by $t=const.$ We define a unit vector field $\BM{n}$, $\BM{n}\cdot\BM{n}=\epsilon=\pm1$.\footnote{Here, for generality, we consider both the cases when $\epsilon=-1$ corresponding to a space-time hypersurface where $\BM{n}$ is timelike, and when $\epsilon=+1$ corresponding to a space-time hypersurface where $\BM{n}$ is spacelike.} The vector field is defined to be stationary, i.e. $\mathsterling_{\ind{\BM{\xi}_{(t)}}}\BM{n}=0$ and hypersurface $\Sigma_{t}$ orthogonal, i.e. $n_{\alpha}\propto\delta_{\alpha}^{\;t}$. Let $\Sigma_{t}$ be spanned by the vectors $e_{(a)}^{\;\alpha}=\delta_{a}^{\;\alpha},\hh n_{\alpha}e_{(a)}^{\;\alpha}=0$, where small Latin letters $(a,b,c,...)$ stand for coordinates on $\Sigma_{t}$, and let $\gamma_{ab}$ be the induced metric on the hypersurface. Then, we can present the space-time metric as
\be\n{2.1}
g^{\alpha\beta}=\epsilon n^{\alpha}n^{\beta}+\gamma^{ab}e_{(a)}^{\alpha}e_{(b)}^{\beta}\,.
\ee
We shall assume that the conditions for Frobenius's theorem hold for the space-time of interest. Namely, using Wald's formulation of Frobenius's theorem \cite{Wald}, we say that for the given space-time (or in a simply connected open subdomain ${\cal D}$) the following conditions hold: 
\begin{itemize}
\item[C1:] $\xi_{(t)}^{[\alpha}\xi_{(\phi)}^{\beta}\nabla^\gamma\xi_{(t)}^{\delta]}$ and $\xi_{(t)}^{[\alpha}\xi_{(\phi)}^{\beta}\nabla^\gamma\xi_{(\phi)}^{\delta]}$vanish at at least one point of the space-time;
\item[C2:] $\xi_{(t)}^{[\alpha}\xi_{(\phi)}^{\beta}R^{\gamma]}_{\,\,\,\,\,\delta}\xi_{(t)}^{\delta}=\xi_{(t)}^{[\alpha}\xi_{(\phi)}^{\beta}R^{\gamma]}_{\,\,\,\,\,\delta}\xi_{(\phi)}^{\delta}=0$\,.
\end{itemize}
These conditions imply that the two-parametric abelian group of isometries $\{\varphi_{t},\varphi_{\phi}\}$ is orthogonally transitive, and thus invertible in ${\cal D}$ \cite{Carter}. In other words, 2-dimensional surfaces of transitivity of the isometry group which are spanned by the Killing vectors $\BM{\xi}_{(t)}$ and $\BM{\xi}_{(\phi)}$ are orthogonal to the family of surfaces of conjugate dimension. As a result, one can present the space-time metric as a direct sum of the metrics on the 2-dimensional orthogonal surfaces (see Eqs. \eq{2.11a}-\eq{2.11c} below).  

One of the cases to satisfy the conditions is to consider a vacuum space-time region, which contains a non-empty subset of fixed points of the group.  Another, less trivial example is the case of electromagnetic space-times, which we consider here. It was showed by Carter \cite{Carter,Carter1} that the conditions hold for a stationary and axisymmetric electromagnetic field. Because the metric tensor is invertible, an addition of a cosmological constant to the Einstein equations doesn't violate the conditions.  

We choose $e_{(\phi)}^{\alpha}=\xi^{\alpha}_{(\phi)}$. Then the Killing vector $\xi^{\alpha}_{(t)}$ lies in a 2-dimensional subspace spanned by $\{\BM{n},\BM{\xi}_{(\phi)}\}$. We define
\be\n{2.2}
\BM{\xi}_{(t)}\cdot\BM{n}=k\,,\hh \BM{\xi}_{(\phi)}\cdot\BM{\xi}_{(\phi)}=\gamma_{\phi\phi}\,,\hh \BM{\xi}_{(t)}\cdot\BM{\xi}_{(\phi)}=\omega\gamma_{\phi\phi}\,,
\ee
where $k$ an $\omega$ are some scalar functions. Then,
\be\n{2.3}
\BM{\xi}_{(t)}=k\BM{n}+\omega\BM{\xi}_{(\phi)}\,.
\ee
In the coordinate basis $(t,x^{a})$,
\be\n{2.4}
n^{\alpha}=k^{-1}(\delta^{\alpha}_{~t}-\omega^{a}\delta^{\alpha}_{~a})\,,\hh \omega^{a}=\omega\delta^{a}_{~\phi}\,,\hh n_{\alpha}=\epsilon k\delta_{\alpha}^{~t}\,,
\ee
and the metric \eq{2.1} takes the following form:
\be\n{2.5}
g^{\alpha\beta}=
\begin{pmatrix}
\epsilon/k^{2} &\hh -\epsilon\omega^{b}/k^{2} \\
-\epsilon\omega^{a}/k^{2} &\hh \gamma^{ab}+\epsilon\omega^{a}\omega^{b}/k^{2}      
\end{pmatrix}\,.
\ee  
The covariant from of the space-time metric $g_{\alpha\beta}$ is
\be\n{2.6}
g_{\alpha\beta}=
\begin{pmatrix}
\epsilon k^{2}+\omega^{c}\omega_{c} &\hh \omega_{a} \\
\omega_{b} &\hh \gamma_{ab}      
\end{pmatrix}\,.
\ee
Here $\gamma_{ac}\gamma^{cb}=\delta_{a}^{\;b}$ and Latin indices of the objects living in $\Sigma_{t}$ are lowered and raised by $\gamma_{ab}$ and $\gamma^{ab}$, respectively, e.g., $\omega_{a}=\gamma_{ab}\omega^{b}$.

To further specify our metric, we assume that $\nabla_\alpha k\nabla^\alpha k$ vanishes nowhere in the domain of interest. Thus, one can take $k$ as one of the space-time coordinates and define $e_{(k)}^{\alpha}=\delta^{\alpha}_{~k}$. We denote by $x$ the remaining spatial coordinate, such that $e_{(x)}^{\alpha}=\delta^{\alpha}_{~x}$. Let us consider a 2-dimensional spacelike surface $\Sigma_{t,k}$ defined by $t,k=const$ and spanned by $\{e_{(x)}^{\alpha},e_{(\phi)}^{\alpha}\}$ with the metric $h_{AB}$ ($x^{A}=(x,\phi)$) on it, which can always be brought to diagonal form. The spacelike vector $\nabla_\alpha k=\delta_\alpha^{~k}$ is orthogonal to such a surface and we define
\ba\n{2.10}
\nabla_\alpha k\nabla^\alpha k=\delta_\alpha^{~k} g^{\alpha\beta} \delta^{~k}_{\beta}=g^{kk}=-\epsilon\kappa^2\,,
\ea
so that for different signs of $\epsilon$ the metric signature is preserved.

As a result, the metric \eq{2.6} can be written in the following form:
\ba
&&ds^2=(\epsilon k^2+\omega^{c}\omega_{c})dt^2+2\omega_{a}dtdx^{a}+\gamma_{ab}dx^{a}dx^{b}\,,\n{2.11a}\\
&&\gamma_{ab}dx^{a}dx^{b}=-\epsilon\kappa^{-2}dk^2+h_{AB}dx^Adx^B\,,\n{2.11b}\\
&&h_{AB}dx^Adx^B=h_{xx}dx^{2}+h_{\phi\phi}d\phi^{2}\,.\n{2.11c}
\ea
The expressions \eq{2.11a}-\eq{2.11c} define a (1+1+2) split of the space-time. We shall use capital Latin letters $(A,B,C,...)$ for the horizon surface coordinates.

\section{Reduction of the curvature tensor} 

In this section we define relations between the Riemann curvature tensor of the 4-dimensional space-time and geometrical quantities of a 2-dimensional surface $\Sigma_{t,k}$. This procedure we shall accomplish in two steps. In the first step, we consider relations between the 4-dimensional Riemann curvature tensor and the intrinsic and extrinsic geometry of a hypersurface $\Sigma_{t}$. Such relations can be found by introducing the projection tensor
\be\n{3.1}
P_{\alpha\beta}=g_{\alpha\beta}-\epsilon n_{\alpha}n_{\beta}\,,
\ee
and using the definition of the Riemann tensor (for details see, e.g., \cite{MTW,St,SKMHH}). The relations are the following:
\ba
&&\hspace{-0.65cm}R^{\alpha}_{\,\,\,a\beta b}n_{\alpha}n^{\beta}={\cal \bar{S}}_{ac}{\cal \bar{S}}^{c}_{\,\,\,b}-k^{-1}(\epsilon k_{|ab}-[\mathsterling_{\ind{\BM{\omega}}}{\cal \bar{S}}]_{ab}+{\cal \bar{S}}_{ab,t})\,,\n{3.2a}\\
&&\hspace{-0.25cm}R^{\alpha}_{\,\,\,abc}n_{\alpha}={\cal \bar{S}}_{ab|c}-{\cal \bar{S}}_{ac|b}\,,\n{3.2b}\\
&&\hspace{0.15cm}R_{abcd}={\bar R}_{abcd}-\epsilon({\cal \bar{S}}_{ac}{\cal \bar{S}}_{bd}-{\cal \bar{S}}_{ad}{\cal \bar{S}}_{bc})\,,\n{3.2c}
\ea
where ${\cal \bar{S}}_{ab}$ is the extrinsic curvature of a hypersurface $\Sigma_{t}$ defined as 
\ba\n{3.5}
{\cal \bar{S}}_{\alpha\beta}&=&{\cal \bar{S}}_{\beta\alpha}\equiv P^{\mu}_{\alpha}P^{\nu}_{\beta}\nabla_{\mu}n_{\nu}\,,\\
{\cal \bar{S}}_{ab}&=&-k^{-1}\omega_{(a|b)}+\frac{1}{2}k^{-1}\gamma_{ab,t}\,,
\ea
(although the last term vanishes, we shall keep it for the second step),   
$\mathsterling_{\ind{\BM{\omega}}}{\cal \bar{S}}$ is the Lie derivative of ${\cal \bar{S}}_{ab}$ in the direction of the vector field $\BM{\omega}$,
\be\n{3.6} 
[\mathsterling_{\ind{\BM{\omega}}}{\cal \bar{S}}]_{ab}={\cal \bar{S}}_{ab,c}\omega^{c}+{\cal \bar{S}}_{cb}\omega^{c}_{\,\,\,,a}+{\cal \bar{S}}_{ac}\omega^{c}_{\,\,\,,b}\,,
\ee
and ${\bar R}_{abcd}$ is the Riemann tensor corresponding to the metric $\gamma_{ab}$. Here and in what follows, the barred geometric quantities correspond to hypersurfaces $\Sigma_{t}$, and the stroke $|$ stands for the covariant derivative defined with respect to the metric $\gamma_{ab}$.

Using the relations \eq{3.2a}-\eq{3.2c}, we derive the following components of the Riemann tensor:
\ba
R^{ta}_{\,\,\,\,\,bc}&=&k^{-1}H^{a}_{\,\,\,bc}\,,\n{3.7a}\\
R^{ab}_{\,\,\,\,\,cd}&=&{\bar R}^{ab}_{\,\,\,\,\,cd}-Q^{ab}_{cd} -2k^{-1}\omega^{[a}H^{b]}_{\,\,\,cd}\,,\n{3.7b}\\
R^{ta}_{\,\,\,\,\,tb}&=&k^{-1}\omega^{c}H^{a}_{\,\,\,cb}-Q^{ac}_{bc}-k^{-1}(k^{|a}_{\,\,\,\,\,|b}-L^{a}_{b})\,,\n{3.7c}\\
R^{bc}_{\,\,\,\,\,ta}&=&\omega^{d}({\bar R}^{bc}_{\,\,\,\,\,da}-Q^{bc}_{da})+2\omega^{[b}Q^{c]d}_{ad}+\epsilon kH_{a}^{\,\,\,bc}\non\\
&-&2k^{-1}(\omega^{d}\omega^{[b}H^{c]}_{\,\,\,da}-\omega^{[b}k^{|c]}_{\,\,\,\,\,|a}+\omega^{[b}L^{c]}_{a})\,,\n{3.7d}
\ea
where
\be
H^{a}_{\,\,\,bc}=2\epsilon{\cal \bar{S}}^{a}_{\,\,\,[b|c]}\,,\hh Q^{ab}_{cd}=2\epsilon{\cal \bar{S}}^{a}_{\,\,\,[c}{\cal \bar{S}}^{b}_{\,\,\,d]}\,,\hh L^{a}_{b}=\epsilon[\mathsterling_{\ind{\BM{\omega}}}{\cal \bar{S}}]^{a}_{b}\,.\n{3.8}
\ee

In the second step, we write the components of the 3-dimensional Riemann tensor ${\bar R}_{abcd}$ in terms of geometric quantities corresponding to a 2-dimensional surface $\Sigma_{t,k}$. Applying the replacements corresponding to $g_{\alpha\beta}\rightarrow\gamma_{ab}$,
\ba\n{3.9}
\epsilon&\rightarrow&-\epsilon,\hh t\rightarrow k,\hh k\rightarrow \kappa^{-1},\hh \omega_{a}\rightarrow 0,\hh \gamma_{ab}\rightarrow h_{AB}\,,\non\\
R^{\alpha\beta}_{\quad\gamma\delta}&\rightarrow&{\bar R}^{ab}_{\quad cd},\hh {\bar R}^{ab}_{\quad cd}\rightarrow {\cal R}^{AB}_{\quad CD},\hh {\cal \bar{S}}_{ab}\rightarrow {\cal S}_{AB}\,,
\ea
to the relations \eq{3.7a}-\eq{3.7d} we derive 
\ba
&&\hspace{-0.65cm}{\bar R}^{AB}_{\,\,\,\,\,\,\,\,CD}={\cal R}^{AB}_{\,\,\,\,\,\,\,\,CD}+\epsilon({\cal S}^{A}_{\,\,\,\,\,C}{\cal S}^{B}_{\,\,\,D}-{\cal S}^{A}_{\,\,\,D}{\cal S}^{B}_{\,\,\,C})\,,\n{3.10a}\\
&&\hspace{-0.65cm}{\bar R}^{kA}_{\,\,\,\,\,BC}=-\epsilon\kappa({\cal S}^{A}_{\,\,\,B;C}-{\cal S}^{A}_{\,\,\,C;B})\,,\n{3.10b}\\
&&\hspace{-0.65cm}{\bar R}^{BC}_{\,\,\,\,\,kA}=\kappa^{-1}({\cal S}^{B\,\,\,;C}_{\,\,\,A}-{\cal S}^{C\,\,\,;B}_{\,\,\,A})\,,\n{3.10c}\\
&&\hspace{-0.65cm}{\bar R}^{kA}_{\,\,\,\,\,kB}=\epsilon\kappa h^{AC}{\cal S}_{CB,k}-\epsilon{\cal S}^{A}_{\,\,\,C}{\cal S}^{C}_{\,\,\,B}-\kappa(\kappa^{-1})^{;A}_{\,\,\,\,\,;B}\,,\n{3.10d}
\ea
where ${\cal R}_{ABCD}$ is the Riemann tensor of a 2-dimensional surface $\Sigma_{t,k}$,
\ba
{\cal R}^{AB}_{\,\,\,\,CD}&=&\frac{1}{2}(\delta^{A}_{\,\,C}\delta^{B}_{~D}-\delta^{A}_{\,\,D}\delta^{B}_{\,\,C}){\cal R}\,,\n{3.10e}\\
{\cal R}^{AB}_{\,\,\,\,CD}{\cal R}^{CD}_{\,\,\,\,AB}&=&{\cal R}^2\,,\hhh{\cal R}^{A}_{\,\,B}=\frac{1}{2}\delta^{A}_{\,\,B}{\cal R}\,,\n{3.10f}
\ea
and ${\cal S}_{AB}$ is its extrinsic curvature,
\be\n{3.11}
{\cal S}_{AB}=\frac{1}{2}\kappa h_{AB,k}\,.
\ee
Here and in what follows, the semicolon stands for the covariant derivative defined with respect to the 2-dimensional metric $h_{AB}$.

To express the other 4-dimensional quantities that enter the expressions \eq{3.7a}-\eq{3.7d} in terms of 2-dimensional ones, we shall use the Christoffel symbols corresponding to the metric $\gamma_{ab}$:
\ba\n{3.12}
{\bar \Gamma}^{k}_{\;kk}&=&-\kappa^{-1}\kappa_{,k},\hh {\bar \Gamma}^{k}_{\;kA}=-\kappa^{-1}\kappa_{,A},\hh
{\bar \Gamma}^{k}_{\;AB}=\epsilon\kappa{\cal S}_{AB}\,,\non\\
\hh {\bar \Gamma}^{A}_{\;kk}&=&-\epsilon\kappa^{-3}\kappa^{,A},\hh {\bar \Gamma}^{A}_{\;kB}=\kappa^{-1}{\cal S}^{A}_{\;B},\hh{\bar \Gamma}^{A}_{\;BC}=\pi^{A}_{\;BC}\,,\non\\
\ea
where $\pi^{A}_{\;BC}$'s are the Christoffel symbols associated with the metric $h_{AB}$,
\ba\n{3.12a}
\pi_{xxx}&=&\frac{1}{2}h_{xx,x}\,,\hh \pi_{x\phi\phi}=-\frac{1}{2}h_{\phi\phi,x}\,,\hh \pi_{\phi x\phi}=\frac{1}{2}h_{\phi\phi,x}\,.\non\\
\ea
Then, for the metric \eq{2.6} we derive
\ba
&&\hspace{-0.65cm}k^{|k}_{\hh|k}=-\epsilon\kappa\kappa_{,k},\hh k^{|k}_{\hh|A}=-\epsilon\kappa\kappa_{,A},\hh
k^{|A}_{\hh|k}=\kappa^{-1}\kappa^{,A}\,,\n{3.13a}\\ 
&&\hspace{-0.65cm}k^{|A}_{\hh|B}=-\epsilon\kappa{\cal S}^{A}_{\hh B},\hh
k^{|a}_{~|a}=-\epsilon\kappa (\kappa_{,k}+{\cal S}),\hh {\cal S}={\cal S}^{A}_{\hh A}\,.\n{3.13b}
\ea
The nonzero extrinsic curvature components read 
\be
{\bar S}_{kA}=-\frac{1}{2}k^{-1}h_{\phi\phi}\,\omega_{,k}\delta^{\phi}_{~A}\,,\hh {\bar S}_{AB}=-k^{-1}h_{\phi\phi}\,\delta^{\phi}_{~(A}\omega_{,B)}\,.\n{3.14}
\ee
Note that because $\phi$ is a Killing coordinate, ${\cal \bar{S}}^{a}_{\,\,\,a}=0$.

The expressions above allow us to present the $4$-dimensional components of the Riemann and Ricci tensors in terms of the $2$-dimensional ones, associated with the metric $h_{AB}$ and the $4$-dimensional metric functions.  

\section{The Einstein equations and curvature invariants}

In this section we construct the Einstein equations corresponding to stationary space-time \eq{2.11a}-\eq{2.11c} with an electromagnetic field and a cosmological constant and derive expressions for scalar curvature invariants. The Einstein equations read
\be\n{4.1}
R^{\alpha}_{\;\beta}=\Lambda\delta^{\alpha}_{\;\beta}+8\pi (T^{\alpha}_{\;\beta}-\frac{1}{2}T\delta^{\alpha}_{\;\beta})\,,\hh T=T^{\alpha}_{\;\alpha}\,.
\ee

\subsection{The electromagnetic field}

Here we shall consider an electromagnetic field without sources in a simply connected space-time domain ${\cal D}$. The electromagnetic stress-energy tensor is
\be\n{4.2e}
T^{\alpha}_{\beta}=\frac{1}{4\pi}(F^{\alpha\gamma}F_{\beta\gamma}-\frac{1}{4}\delta^{\alpha}_{\,\,\,\beta}F^{2})\,,\hh F^{2}=F_{\alpha\beta}F^{\alpha\beta}\,,
\ee
and $T=0$. The electromagnetic field tensor $F_{\alpha\beta}$ can be derived from a 4-vector potential $\BM{A}$ which will be assumed to satisfy the group invariance conditions
\be\n{4.2a}
\mathsterling_{\ind{\BM{\xi}_{(t)}}}\BM{A}=0\,,\hhh \mathsterling_{\ind{\BM{\xi}_{(\phi)}}}\BM{A}=0\,, 
\ee
and the electromagnetic potential circularity condition \cite{Carter1},
\be
A_{[\alpha}\xi_{(t)\beta}\xi_{(\phi)\gamma]}=0\,.
\ee
As a result, it depends only on the $k$ and $x$ coordinates and can be presented in the form
\be\n{4.2b}
A_{\alpha}=-\Phi\delta_{\alpha}^{\,\,\,t}+{\cal A}\delta_{\alpha}^{~\phi}\,, 
\ee
where $\Phi=\Phi(k,x)$ and ${\cal A}={\cal A}(k,x)$.  The corresponding electromagnetic field tensor $F_{\alpha\beta}=A_{\beta,\alpha}-A_{\alpha,\beta}$ has the following components:
\ba\n{4.2c}
F_{ta}&=&-F_{at}=\Phi_{,a}\,,\hhh F_{ab}=2{\cal A}_{[,a}\delta_{b]}^{\,\,\,\phi}\,,\non\\
F^{ta}&=&-F^{at}=\epsilon k^{-2}(\Phi^{,a}+\omega {\cal A}^{,a})\,,\\
F^{ab}&=&2(h^{\phi\phi}{\cal A}^{[,a}+\omega F^{t[a})\delta^{b]}_{\,\,\,\phi}\,.\non
\ea
The Maxwell equations for a source-free electromagnetic field read
\be\n{4.2d}
\nabla_{\beta}F^{\alpha\beta}=\frac{1}{\sqrt{-g}}(\sqrt{-g}F^{\alpha\beta})_{,\beta}=0\,,
\ee
where $g=\text{det}(g_{\alpha\beta})=-k^{2}\kappa^{-2}h$ and $h=\text{det}(h_{AB})$. Using the expressions \eq{4.2c} the Maxwell equations can be written in the form
\ba
&&[k^{-1}\kappa^{-1}\sqrt{h}\,(\Phi^{,a}+\omega\,{\cal A}^{,a})]_{,a}=0\,,\n{M1}\\
&&k^{-1}\kappa^{-1}\sqrt{h}\,\omega_{,a}(\Phi^{,a}+\omega\,{\cal A}^{,a})+\epsilon[k\kappa^{-1}\sqrt{h}\,h^{\phi\phi}\,{\cal A}^{,a}]_{,a}=0\,.\non
\ea 
The electromagnetic field invariant and energy density are the following:
\ba
F^{2}&=&2{\cal A}_{,a}{\cal A}^{,a}h^{\phi\phi}+2\epsilon k^{-2}(\Phi_{,a}+\omega {\cal A}_{,a})(\Phi^{,a}+\omega {\cal A}^{,a})\,,\non\\
{\cal E}&=&\frac{\epsilon}{16\pi}(F^{2}-4{\cal A}_{,a}{\cal A}^{,a}h^{\phi\phi})\,.\n{4.2f}
\ea

\subsection{The Einstein equations}

The Ricci tensor components and the Ricci scalar read 
\ba
R^{t}_{\,\,\,t}=R^{ta}_{\,\,\,\,\,ta},\hh R^{t}_{\,\,\,a}=R^{tb}_{\,\,\,\,\,ab},\hh R^{a}_{\,\,\,t}=R^{ab}_{\,\,\,\,\,tb}\,,\non\\
R^{a}_{\,\,\,b}=R^{ta}_{\,\,\,\,\,tb}+R^{ac}_{\,\,\,\,\,\,bc},\hh R=2R^{ta}_{\,\,\,\,\,ta}+R^{ab}_{\,\,\,\,\,\,ab}\,.\n{4.2}
\ea
With the aid of the expressions \eq{3.7a}-\eq{3.7d} and \eq{4.2}, the Einstein equations \eq{4.1} can be written as follows:
\ba
&&k^{-1}H^{b}_{\,\,\,ab}=8\pi T^{t}_{\,\,\,a}\,,\hh W^{ab}_{ab}=2\Lambda+2\tilde{T}^{a}_{\,\,\,a}\,,\n{4.3a}\\
&&V^{a}_{~b}-Q^{ac}_{bc}-k^{-1}(k^{|a}_{\,\,\,\,\,|b}-L^{a}_{b})=0\,,\n{4.3b}
\ea
where
\ba\n{4.3b1}
W^{ab}_{cd}&=&{\bar R}^{ab}_{\,\,\,\,\,cd}-Q^{ab}_{cd}\,,\hh\tilde{T}^{a}_{\,\,\,b}=8\pi(T^{a}_{\,\,\,b}+\omega^{a}T^{t}_{\,\,\,b})\,,\non\\
V^{a}_{~b}&=&W^{ac}_{bc}-\tilde{T}^{a}_{\,\,\,b}-\Lambda\delta^{a}_{\,\,\,b}\,.
\ea

\subsection{The scalar curvature invariants}

As we mentioned in Introduction, in this paper we consider the Kretschmann, Chern-Pontryagin, and Euler curvature invariants defined as follows:
\ba
{\cal K}_{1}&=&R_{\alpha\beta\gamma\delta}R^{\alpha\beta\gamma\delta}={\cal C}_{\alpha\beta\gamma\delta}{\cal C}^{\alpha\beta\gamma\delta}+2R_{\alpha\beta}R^{\alpha\beta}-\frac{1}{3}R^{2}\,,\non\\
{\cal K}_{2}&=&^{*\!}R_{\alpha\beta\gamma\delta}R^{\alpha\beta\gamma\delta}=^{*\!\!}{\cal C}_{\alpha\beta\gamma\delta}{\cal C}^{\alpha\beta\gamma\delta}\,,\n{I}\\
{\cal K}_{3}&=&^{*\!}R^{*}_{\alpha\beta\gamma\delta}R^{\alpha\beta\gamma\delta}=-{\cal C}_{\alpha\beta\gamma\delta}{\cal C}^{\alpha\beta\gamma\delta}+2R_{\alpha\beta}R^{\alpha\beta}-\frac{2}{3}R^{2}\,,\non
\ea
respectively. Here ${\cal C}_{\alpha\beta\gamma\delta}$ is the Weyl tensor, and the star symbol stands for the left and the right Hodge dual quantities, e.g.,
\be
^{*\!}R_{\alpha\beta\gamma\delta}=\frac{1}{2}\varepsilon_{\alpha\beta\mu\nu}R^{\mu\nu}_{\,\,\,\,\,\,\gamma\delta}\,,\hh R^{*}_{\alpha\beta\gamma\delta}=\frac{1}{2}\varepsilon_{\gamma\delta\mu\nu}R_{\alpha\beta}^{\,\,\,\,\,\,\,\mu\nu}\,.
\ee
Here
\ba\n{LC4}
\varepsilon_{\alpha\beta\gamma\delta}&=&\sqrt{-g}\,\bar{\varepsilon}_{\alpha\beta\gamma\delta}\,,\hh \varepsilon^{\alpha\beta\gamma\delta}=\frac{\bar{\varepsilon}^{\alpha\beta\gamma\delta}}{\sqrt{-g}}\,,\non\\
\bar{\varepsilon}_{tkx\phi}&=&+1\,,\hh \bar{\varepsilon}^{tkx\phi}=-1\,,
\ea
is 4-dimensional Levi-Civita pseudo tensor. 

Using the expressions of this section and the Riemann tensor components \eq{3.7a}-\eq{3.7d} we can write the Kretschmann and  Chern-Pontryagin invariants in the form
\ba\n{4.4a}
{\cal K}_{1}&=&R^{ab}_{\,\,\,\,\,\,cd}R^{cd}_{\,\,\,\,\,\,ab}+4R^{ta}_{\,\,\,\,\,\,bc}R^{bc}_{\,\,\,\,\,\,ta}+4R^{ta}_{\,\,\,\,\,\,tb}R^{tb}_{\,\,\,\,\,\,ta}\,\non\\
&=&W^{ab}_{cd}W^{cd}_{ab}+4V^{a}_{~b}V^{b}_{~a}+4\epsilon\,H^{a}_{\,\,\,bc}H_{a}^{\,\,bc}\,,\n{4.4a}\\
{\cal K}_{2}&=&^{*\!}R^{ab}_{\,\,\,\,\,\,cd}R^{cd}_{\,\,\,\,\,\,ab}+2(^{*\!}R^{ta}_{\,\,\,\,\,\,bc}R^{bc}_{\,\,\,\,\,\,ta}+^{*\!}R^{bc}_{\,\,\,\,\,\,ta}R^{ta}_{\,\,\,\,\,\,bc})\non\\
&+&4^{*\!}R^{ta}_{\,\,\,\,\,\,tb}R^{tb}_{\,\,\,\,\,\,ta}=2\epsilon_{a}^{~bc}(H^{a}_{~de}W^{de}_{bc}-2H^{d}_{~bc}V^{a}_{~d})\,.\n{4.4b}
\ea
Here
\ba\n{LC3}
\epsilon_{abc}&=&\sqrt{|\gamma|}\,\bar{\epsilon}_{abc}\,,\hh \epsilon^{abc}=\frac{\bar{\epsilon}^{abc}}{\sqrt{|\gamma|}}\,,\non\\
\gamma&=&\text{det}(\gamma_{ab})=-\epsilon\kappa^{-2}h\,,\\
\bar{\epsilon}_{kx\phi}&=&+1\,,\hh \bar{\epsilon}^{kx\phi}=-\epsilon\,,\non
\ea
is 3-dimensional Levi-Civita pseudo tensor. Note that according to the definition of $H^{a}_{~bc}$ [see \eq{3.8}], we have $\epsilon^{abc}H_{abc}=0$.
 
The Euler curvature invariant can be derived from the Kretschmann invariant,  the square of the Ricci tensor $R_{\alpha\beta}R^{\alpha\beta}$, and the Ricci scalar $R$,
\ba
&&R^{\alpha}_{~\beta}R^{\beta}_{~\alpha}=4\Lambda^{2}+64\pi^{2}T^{\alpha}_{~\beta}T^{\beta}_{~\alpha}\,,\n{4.4c}\\
&&R=R^{\alpha}_{~\alpha}=4\Lambda\,,\n{4.4d}
\ea
through the following expression:
\be\n{4.5}
{\cal K}_{3}=4R_{\alpha\beta}R^{\alpha\beta}-R^{2}-{\cal K}_{1}\,.
\ee

\section{Geometric properties of the Killing Horizon}

\subsection{Killing Horizon}

Let us consider the Killing vector field
\be\n{2.7}
\BM{\chi}=\BM{\xi}_{(t)}+\Omega\BM{\xi}_{(\phi)}\,,
\ee
where $\Omega=const.$ We have
\be\n{2.8}
\BM{\chi}\cdot\BM{\chi}=\epsilon k^{2}+(\omega^{2}+2\omega\Omega+\Omega^{2})\gamma_{\phi\phi}\,,
\ee
and the condition 
\be\n{2.9}
\omega\circeq-\Omega\,,
\ee
implies that $k=0$ is a Killing horizon, i.e., $\BM{\chi}\cdot\BM{\chi}\circeq0$. According to this condition, $\BM{\chi}$ is hypersurface orthogonal on $k=0$, i.e., $\chi_{[\alpha}\nabla_{\beta}\chi_{\gamma]}\circeq0$. A meaning of the condition \eq{2.9} can be seen from the definition of the angular velocity of a horizon,
\be\n{2.9a}
\Omega^{{\cal H}}\circeq-\frac{g_{t\phi}}{g_{\phi\phi}}=\Omega\,,
\ee
which implies that the Killing horizon rotates as though it were a solid body, i.e. the condition \eq{2.9} implies rigidity of the Killing horizon. 

The metric function $\kappa$ calculated on the Killing horizon coincides with its surface gravity, 
\be\n{5.10}
\kappa^{2}\circeq\frac{\epsilon}{2}\lim_{k\to0}(\nabla_{\alpha}\chi_{\beta})(\nabla^{\alpha}\chi^{\beta})\,.
\ee
If $\kappa$ vanishes on the Killing horizon, it is called {\em degenerate} (or extremal), otherwise, it is called {\em nondegenerate} (or non-extremal). In the following calculations we shall assume that $\kappa\ne0$.

The Killing horizon is a totally geodesic hypersurface \cite{Boyer}, which implies its extrinsic curvature vanishes \cite{Eisenhart}. To calculate the extrinsic curvature of a hypersurface $\Sigma_{k}$ ($k=const$) we define a unit vector $N_{\alpha}$ orthogonal to it,
\be\n{5.1}
N_{\alpha}=-\epsilon\kappa^{-1}\delta_{\alpha}^{~k}\,,\hhh N^{\alpha}=\kappa\,\delta^{\alpha}_{~k}\,,\hhh N^{\alpha}N_{\alpha}=-\epsilon\,,
\ee
and the corresponding projection tensor,
\be\n{5.2}
\Pi_{\alpha\beta}=g_{\alpha\beta}+\epsilon\,N_{\alpha}N_{\beta}\,.
\ee
The extrinsic curvature of a hypersurface $\Sigma_{k}$ is defined as
\be\n{5.3}
{\cal \tilde{S}}_{\alpha\beta}={\cal \tilde{S}}_{\beta\alpha}\equiv \Pi^{\mu}_{\alpha}\Pi^{\nu}_{\beta}\nabla_{\mu}N_{\nu}\,,
\ee
and its nonzero components read
\ba
{\cal \tilde{S}}_{tt}&=&\epsilon\kappa k+\kappa\omega h_{\phi\phi}\omega_{,k}+\frac{1}{2}\kappa\omega^{2}h_{\phi\phi,k}\,,\n{5.4}\\
{\cal \tilde{S}}_{tA}&=&\frac{1}{2}\kappa(h_{\phi\phi}\omega)_{,k}\delta^{\phi}_{~A}\,,\hhh {\cal \tilde{S}}_{AB}={\cal S}_{AB}=\frac{1}{2}\kappa h_{AB,k}\,.\non
\ea
Thus, for a nondegenerate Killing horizon we have
\be\n{5.5}
\omega_{,k}\circeq0\,\hhh h_{AB,k}\circeq0\,.
\ee
 Geometric and field invariants are finite on a regular Killing horizon. In particular, the invariants $Q^{ab}_{ab}$ and $F^{2}$ are finite on $k=0$. Thus, according to the expressions \eq{3.14} [see \eq{2.9} as well] and the Maxwell equations \eq{M1}, we have
\be\n{5.6}
\omega_{,A}\circeq0\,,\hh \Phi_{,k}\circeq0\,,\hh{\cal A}_{,k}\circeq0\,,\hh \Phi_{,A}+\omega {\cal A}_{,A}\circeq0\,.
\ee  
We consider the metric and the field functions $\varphi(k,x)=\{\omega\,,\kappa\,,h_{AB}\,,\Phi+\omega {\cal A}\}$ on and at the vicinity of the Killing horizon of class $C^{r}$, $r\geq2$ in our coordinates. Then according to the Schwarz' (Clairaut's) theorem,   
\ba\n{5.7}
&&\lim_{k\to0}k^{-1}\varphi_{,A}=\varphi_{,Ak}(0,x^{A})=\varphi_{,kA}(0,x^{A})\non\\
&&=\lim_{\Delta x^{A}\to0}\frac{\varphi_{,k}(0,x^{A}+\Delta x^{A})-\varphi_{,k}(0,x^{A})}{\Delta x^{A}}=0\,,
\ea  
where the last equality follows from \eq{5.5} and \eq{5.6}. Using these conditions and taking the limit $k\to0$ in the expression \eq{3.14} one can show that
\be\n{5.8}
{\bar S}_{AB}\circeq0\,,
\ee
and ${\bar S}_{kA}$ is finite on the horizon. As a result, the Lie derivative of ${\bar S}_{ab}$ \eq{3.6} vanishes on the horizon. Then, using the Einstein equations \eq{4.3a}-\eq{4.3b} one can see that $k^{|a}_{\,\,\,\,\,|b}$ vanishes on the horizon and the expressions \eq{3.13a}-\eq{3.13b} give 
\be\n{5.9}
\kappa_{,k}\circeq0\,,\hhh \kappa_{,A}\circeq0\,.
\ee
Thus, the quantities $\omega$, $\kappa$, and $\Phi+\omega {\cal A}$ are constant on the Killing horizon. This result is well-known. It can be derived by using geometric properties of Killing horizons derived in \cite{Boyer} and \cite{Wald2} (see \cite{FN}). The derivation presented here includes the $k$-derivatives of the functions which are used in the derivation of our main results. 

\subsection{Curvature invariants on the Killing Horizon}

In this subsection we derive the relations between the space-time curvature invariants calculated on the Killing horizon. Using the results of the previous subsection and the expressions \eq{3.10a}-\eq{3.10f} we derive 
\ba
W^{kA}_{kB}&\circeq&M^{A}_{B}+M\delta^{A}_{\,\phi}\delta^{\phi}_{\,B}-\frac{1}{4}\delta^{A}_{\,\,B}({\cal R}+F^{2}-2\Lambda),\non\\
W^{kA}_{BC}&\circeq&0\,,\hh W^{BC}_{kA}\circeq 0\,,\hh W^{AB}_{CD}\circeq\frac{{\cal R}}{2}(\delta^{A}_{\,\,C}\delta^{B}_{\,\,D}-\delta^{A}_{\,\,D}\delta^{B}_{\,\,C})\,,\non\\
\tilde{T}^{k}_{k}&\circeq&\frac{1}{2}F^{2}-2M\,,\hh \tilde{T}^{k}_{A}\circeq 0\,,\hh \tilde{T}^{A}_{k}\circeq 0\,,\n{5.10}\\
\tilde{T}^{A}_{B}&\circeq&2M^{A}_{B}+2M\delta^{A}_{~\phi}\delta^{\phi}_{~B}-\frac{1}{2}F^{2}\delta^{A}_{\,\,B}\,,\non
\ea
where
\be\n{5.12b}
M^{A}_{B}=h^{\phi\phi}{\cal A}^{,A}{\cal A}_{,B}\,,\hh M=M^{A}_{A}\,.
\ee
Using this result  we derive the following expressions of the curvature invariants on the Killing horizon:
\ba\n{5.12c}
&&{\cal K}_{1}\circeq3\left({\cal R}+\epsilon\tilde{\mathcal{E}}-\frac{2}{3}\Lambda\right)^2+4\epsilon H^{abc}H_{\,\,\,abc}+2\tilde{{\cal E}}^{2}+\frac{8}{3}\Lambda^{2}\,,\non\\
&&{\cal K}_{2}\circeq6\epsilon^{aAB}H_{aAB}\left({\cal R}+\epsilon\tilde{\mathcal{E}}-\frac{2}{3}\Lambda\right)\,,\non\\
&&{\cal K}_{3}\circeq-3\left({\cal R}+\epsilon\tilde{\mathcal{E}}-\frac{2}{3}\Lambda\right)^2-4\epsilon H^{a}_{\,\,bc}H_{a}^{\,\,bc}+2\tilde{{\cal E}}^{2}-\frac{8}{3}\Lambda^{2}\,,\non\\
&&R_{\alpha\beta}R^{\alpha\beta}\circeq \tilde{{\cal E}}^{2}+4\Lambda^{2}\,,\hh R=4\Lambda\,,\hh \tilde{\cal E}=16\pi{\cal E}\,.
\ea
The factor $\epsilon=\pm1$, which enters the expressions, suggests a discontinuity in the space-time curvature invariants in the case when the Killing horizon separates the space-time into the regions where the vector $\BM{n}$ is timelike and spacelike. However, such a discontinuity is not present, for there is another such factor ``hidden'' in the stationary term $H^{abc}H_{\,\,\,abc}$, so that effectively one has $\epsilon^{2}=1$. 

The expressions \eq{5.12c} are the main result of our paper. The expressions in the preceding subsection were derived for a nondegenerate Killing horizon. However, assuming that the space-time admits the limit of $\kappa\to0$, which can be accomplished by the corresponding limit of the space-time parameters, the final result \eq{5.12c} remains valid for a degenerate Killing horizon as well. The derived expressions generalize the curvature invariants constructed in \cite{AS} for a static Killing horizon to the stationary one.\footnote{In order to compare the expressions, the electromagnetic field invariant given in the paper \cite{AS} has to be rescaled as follows: $F^{2}\to4F^{2}$.}  

\section{Discussion}

Let us summarize our results. We studied the geometric properties of stationary and axisymetric Killing horizons. Such horizons have zero extrinsic curvature, constant surface gravity, angular velocity, and electromagnetic field (the combination $\Phi+\omega{\cal A}$) and the derivatives of these quantities (except for the extrinsic curvature) in the direction orthogonal to the horizon surface vanish.  We derived the relations between the Kretschmann, Chern-Pontryagin, and Euler space-time curvature invariants, as well as the square of the Ricci tensor and the Ricci scalar,  calculated on a Killing horizon in terms of the geometric quantities corresponding to the horizons surface. These relations are generalizations of the analogous known relations for horizons of static 4-dimensional electrovacuum space-times [see \eq{I1}].

There is a direct analogy between the electromagnetic field tensor $F_{\alpha\beta}$ and the Weyl tensor ${\cal C}_{\alpha\beta\gamma\delta}$. Namely, there are the {\em gravitoelectric} and {\em gravitomagnetic} parts of the Weyl tensor (see, e.g. \cite{SKMHH,PT,TPM,Mat}) which we define as follows:
\be
\mathscr{E}_{\alpha\beta}={\cal C}_{\alpha\gamma\beta\delta}u^{\gamma}u^{\delta}\,,\hh \mathscr{B}_{\alpha\beta}=^{*\!\!}{\cal C}_{\alpha\gamma\beta\delta}u^{\gamma}u^{\delta}\,,
\ee
where $u^{\alpha}=-\epsilon n^{\alpha}$ [cf. \eq{2.4}] is the FIDO 4-velocity (see, e.g., \cite{PT,TPM}). Because these fields are orthogonal to $\BM{n}$, they live on a hypersurface $\Sigma_{t}$ and are effectively 3-dimensional tensor fields. According to the symmetries of the Weyl tensor, they are symmetric and traceless. 
As a result, the Weyl invariants ${\cal C}_{\alpha\beta\gamma\delta}C^{\alpha\beta\gamma\delta}$ and $^{*\!}{\cal C}_{\alpha\beta\gamma\delta}C^{\alpha\beta\gamma\delta}$ are analogous to the electromagnetic field invariants, $F_{\alpha\beta}F^{\alpha\beta}=2(\BM{B}^{2}-\BM{E}^{2})$ and $^{*\!}F_{\alpha\beta}F^{\alpha\beta}=4\BM{E}\cdot\BM{B}$, where $\BM{E}$ and $\BM{B}$ are electric and magnetic fields, respectively. 

One can evaluate the gravitoelectric field component which is orthogonal to the horizon surface,
\be
\mathscr{E}^{k}_{k}\circeq\frac{\epsilon}{2}\left({\cal R}+\epsilon\tilde{\mathcal{E}}-\frac{2}{3}\Lambda\right)\,.\n{Hartle}
\ee
This expression is a generalization of Hartle's curvature formula, which was derived by using the Newman-Penrose formalism (see, e.g., \cite{Hartle},\cite{TPM}). It is interesting to note that there is an additional additive contribution (not only through the space-time metric) to the scalar curvature of the horizon surface from the electromagnetic field energy density and the $\Lambda$ term.
We can express the Weyl invariants in therms of $\mathscr{E}^{ab}$ and $\mathscr{B}_{ab}$ as follows:
\ba
{\cal C}_{\alpha\beta\gamma\delta}C^{\alpha\beta\gamma\delta}&=&8({\cal E}_{ab}{\cal E}^{ab}-{\cal B}_{ab}{\cal B}^{ab})\,,\\
^{*\!}{\cal C}_{\alpha\beta\gamma\delta}C^{\alpha\beta\gamma\delta}&=&16\,{\cal E}_{ab}{\cal B}^{ab}\,.
\ea
A comparison with the expressions \eq{5.12c} implies
\ba
{\cal E}_{ab}{\cal E}^{ab}&\circeq&\frac{3}{8}\left({\cal R}+\epsilon\tilde{\mathcal{E}}-\frac{2}{3}\Lambda\right)^2\,,\n{GE}\\
{\cal B}_{ab}{\cal B}^{ab}&\circeq&-\frac{\epsilon}{2}H^{a}_{\,\,bc}H_{a}^{\,\,bc}\,,\n{GM}\\
{\cal E}_{ab}{\cal B}^{ab}&\circeq&\frac{3}{8}\epsilon^{aAB}H_{aAB}\left({\cal R}+\epsilon\tilde{\mathcal{E}}-\frac{2}{3}\Lambda\right)\,.\n{GEM}
\ea
 The gravitomagnetic part \eq{GM}, which is analogous to the electromagnetic expression $\BM{B}^{2}=(\nabla\times\BM{A})^{2}$, is due to the extrinsic curvature ${\cal \bar{S}}_{ab}$ of a hypersurface $\Sigma_{t}$, which, in turn, is analogous to the vector potential $\BM{A}$. The curvature occurs due to the {\em twist} metric function $\omega$. Such a twist gives an additional contribution to the space-time curvature on the Killing horizon.

The geometric properties of the horizons presented here can be used for calculation of space-time curvature at a Killing horizon of 4-dimensional, stationary and axisymmetric electromagnetic space-time with a cosmological constant. The result may be important for applications to holographic models and, more general for understanding of properties of space-time horizons in general. 

\begin{acknowledgments}

The author is grateful to the Natural Sciences and Engineering Research Council of Canada for the financial support and to Professor Don N. Page for reading the manuscript and useful suggestions.

\end{acknowledgments}

\end{document}